\documentclass[
  journal=largetwo,
  manuscript=article-type,
  year=2024,
]{cup-journal}

\usepackage{amsmath}
\usepackage[nopatch]{microtype}
\usepackage{booktabs}

\usepackage{graphicx}	
\usepackage[english]{babel}
\usepackage[utf8]{inputenc}
\usepackage{amssymb}

\graphicspath{{./}{Picture/}}

\title{The global structure of astrospheres: effect of Knudsen number}

\author{S.D. Korolkov}
\affiliation{Space Research Institute (IKI) of Russian Academy of Sciences, Moscow, Russia}
\alsoaffiliation{Lomonosov Moscow State University, Moscow center for fundamental and applied mathematics, Moscow, Russia}
\alsoaffiliation{HSE University, 20 Myasnitskaya Ulitsa, Moscow 101000, Russia}
\email[S.D. Korolkov]{korolkov.msu@mail.ru}

\author{V.V. Izmodenov}
\affiliation{Space Research Institute (IKI) of Russian Academy of Sciences, Moscow, Russia}
\alsoaffiliation{Lomonosov Moscow State University, Moscow center for fundamental and applied mathematics, Moscow, Russia}
\alsoaffiliation{HSE University, 20 Myasnitskaya Ulitsa, Moscow 101000, Russia}


\begin{document}

\begin{abstract}
The interaction between stellar winds and the partially ionized local interstellar medium (LISM) is quite common in astrophysics. However, the main difficulty in describing the neutral components lies in the fact that the mean free path of an interstellar atom, l, can be comparable to the characteristic size of an astrosphere, L (i.e., the Knudsen number, which is equal to l/L, is approximately equal to 1, as in the case of the heliosphere). In such cases, a single-fluid approximation becomes invalid, and a kinetic description must be used for the neutral component.

In this study, we consider a general astrosphere and use a kinetic-gas dynamics model to investigate how the global structure of the astrosphere depends on the Knudsen number. We present numerical results covering an extremely wide range of Knudsen numbers (from 0.0001 to 100). Additionally, we explore the applicability of single-fluid approaches for modeling astrospheres of various sizes. We have excluded the influence of interstellar and stellar magnetic fields in our model to make parametric study of the kinetic effects feasible.

The main conclusion of this work is that, for large astrospheres (with a distance to the bow shock greater than 600 AU) a heated rarefied plasma layer forms in the outer shock layer near the astropause. The formation of this layer is linked to localized heating of the plasma by atoms (specifically, ENAs) that undergo charge exchange again behind the astropause. This process significantly alters the flow structure in the outer shock layer and the location of the bow shock, and it cannot be described by a single-fluid model. Additionally, this paper discusses how atoms weaken the bow shocks at near-heliospheric conditions.
\end{abstract}

\section{Introduction}
\label{Intro}
\begin{figure*}
	\includegraphics[width=0.9\hsize]{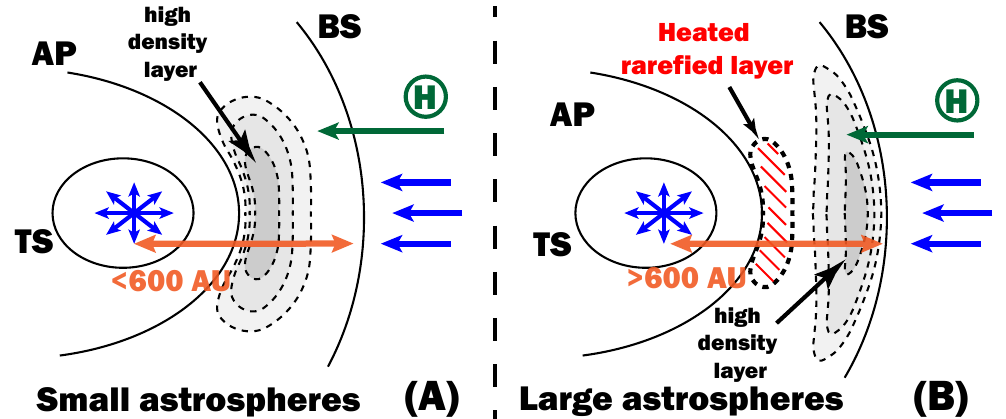}
	\caption{Schematic picture of the interaction of stellar wind with a partially ionized supersonic flow for (A) small and (B) large astrospheres}
	\label{shem1}
\end{figure*}

The problem of the interaction between the solar wind and the interstellar medium has been of particular interest since 1970s. This interest was mainly sparked by experiments on scattered Lyman-alpha radiation in near-solar space, which confirmed the penetration of hydrogen atom fluxes from the local interstellar medium into the hypersonic solar wind (see, for example, \cite{Bertaux1971}, \cite{Thomas1971}). During the same period, spacecraft such as Voyager-1, Voyager-2, Pioneer-10, and Pioneer-11 were launched, providing a significant amount of experimental data on the parameters of the solar wind not only in the Earth orbit but also in the distant heliosphere. With an increase in the quantity and quality of experimental data due to an increasing number of space missions, there is a growing demand for theoretical models. Such models are necessary for a general understanding of the physics of the processes, for predicting various phenomena that may not yet be discovered, and for determining parameters that cannot be measured directly.

The first model of supersonic flow around the solar wind was proposed by \cite{Baranov1970}, assuming a gas-dynamic description of plasma as a completely ionized medium and not taking into account the influence of neutral atoms from the interstellar medium. Later studies showed that neutral atoms have a significant impact on plasma due to resonant charge exchange (see \cite{Wallis1975}), resulting in an exchange of momentum and energy between neutral and charged components. Numerical simulations confirmed this effect (\cite{Baranov1979, Baranov1982, Baranov1979R}), but they assumed a simplified gas-dynamic description of the neutral component. The mean free path of a hydrogen atom is comparable to the size of the heliosphere, so the neutral component must be described kinetically with the solution of the integro-differential Boltzmann equation for the distribution function. \cite{BaranovLebedev1991} proposed a self-consistent model with the solution of the kinetic equation for hydrogen atoms and gas-dynamic equations for the solar wind, ensuring consistency by the method of global iterations. However, only the first step of the iterative algorithm was carried out in this work. A completely self-consistent solution was obtained later \citep{BaranovMalama1993}.

Recently, there has been a lot of interest in the astrospheres of other stars. A wide range of observational data (for example, \cite{Kobul_2016}) makes it possible to determine the position of the discontinuity surfaces (primarily the astropause) and their distance from the parent star, which can be used in numerical models by solving the inverse problem and finding unknown parameters of the star or surrounding medium. 

In this work, we conduct a parametric study of astrosphere structure varying the Knudsen number, $\mathrm{Kn}_\infty$. $\mathrm{Kn}_\infty$ is the ratio of the mean free path of hydrogen atoms (in the undisturbed interstellar medium) and the characteristic size of the astrosphere. The mean free path of hydrogen atoms is determined by particle interactions and depends on the speed of the atom and the properties of the local medium, to a greater extent on the proton number density, to a lesser extent on proton speed and temperature. The size of the astrosphere is determined by stellar wind velocity and the rate of the star mass loss and the parameters of the interstellar medium. Obviously, the lower the Knudsen number, the more efficient the process of charge exchange. In this work, we show that in some cases the charge exchange is capable of reducing the distance from star to the astropause by approximately 50\% compared with one without taking into account atoms (purely gas-dynamic case). 

In addition to practical interest, this work allows us to evaluate the limits of the applicability of the single-fluid approach. 
The single-fluid approach considered in this study is the limiting case of either infinitely large mean free paths, corresponding to pure gas dynamics, or infinitely small mean free paths, which can be described as an effective gas, as discussed in Section~\ref{Lim}. As it turns out, the limits of the single-fluid approach applicability are quite narrow. Besides the differences in the positions of the main discontinuities, the single-fluid approach gives an incorrect description of the structure of the outer astroshethes for large astrospheres. Figure~\ref{shem1} schematically demonstrates this effect. Panel (A) illustrates the structure of an outer shock layer for a small astrospheres (with a distance to the bow shock of less than 600 AU). In this case, the effect of charge exchange causes the maximum of the plasma density to be shifted away from the astropause due to weak local heating, slightly. Note that in the purely gas-dynamic case (without the influence of atoms), as in the case of the single-fluid approach, this maximum is located strictly at the astropause. In the case of a large astrosphere ($>600$ AU, panel B), heating of the outer shock layer by ENAs leads to the formation of a heated rarefied layer of plasma behind the astropause, and the maximum number density in the outer astrosheath occurs near the bow shock (see details in Subsection~\ref{PDL}).


An alternative approach to describe partly ionized plasma flow that has been applied for the heliosphere  is multi-fluid modeling (see, e.g., \cite{1995Pauls, McNutt, 1998Wang, 2000Fahr, 2004Florinski, 2024Bera}, and others). In this approach neutral component is considered as a mixture of several ideal gases. Therefore, the Euler equations for these gases are solved together with MHD equations for the plasma component. The multi-fluid approach is some-time considered as an "intermediate" approach between single-fluid hydrodynamic models and kinetic models. The main argumentation to use the multi-fluid models is their small computational costs as compared with kinetic approach. Also, for some set of model parameters the multi-fluid approaches produce the plasma and atoms distributions are quite close to those obtained in the kinetic models (see, \cite{2005Alexashov, 2006Heerikhuisen, 2011Alouani-Bibi}). Nevertheless, the multi-fluid approach has no any theoretical justification and there are examples (see \cite{Baranov1998,2005Alexashov, 2008Muller}) when using multi-fluid approaches produce physically unreasonable results. In this work, we will not consider multi-fluid approaches and will focus on the kinetic description of hydrogen dynamics. Additionally, we consider the two limiting cases, which are discussed in  Section~\ref{Lim}.

In paper \cite{2024Korolkov}, we investigated the effect of charge exchange on the structure of a single astrospheric shock layer as simply as possible. We considered hydrogen to be constant in the layer and the tangential discontinuity (the astropause) to be planar. These limitations allowed us to conduct research within a fairly narrow range of Knudsen numbers, studying the influence of atoms on the shock layer. However, these limitations do not allow us to quantify accurately the size of either the inner or outer shock layers in the astrosphere. Despite these limitations, we discovered and explained the effect of formation of a heated rarefied layer of plasma even using this simple model. This work continues the previous study by modeling global astrospheres within a wide range of Knudsen numbers (between $10^{-4}$ and $10^2$) while considering the dynamics of hydrogen consistently with the plasma, allowing us to estimate the size of both the shock layers and explore the structure of astrospheres.

We restrict this study to purely unmagnetized models, neglecting any effects of interstellar or stellar magnetic fields. This allows us to reduce the parameter space of the problem and make feasible to attain the main goal of our work that is to quantify kinetic effects on the spatial extent and large-scale properties of the astrosphere. In addition, the magnetic fields of stars and their surrounding environments are largely unknown, and their estimates are wide-ranging. However, it is well-established that magnetic fields can significantly affect the structure of astrospheres, changing their shape qualitatively (for example, see tube-like astrospheres in \cite{opher15, Korolkov_2021}). For the heliosphere, the influence of magnetic fields on the large-scale structures is significant and leads to several effects, such as asymmetry of the termination shock and the heliopause, the deviation of interstellar hydrogen atom inflow, magnetic reconnection and others. These effects have been studied in several papers, including those by \cite{2005Izmodenov, 2009Pogorelov, 2011Alouani-Bibi, izmodenov_alexashov2015}. Additionally, there is a series of papers by \cite{ser2004Pogorelov, ser2006Pogorelov, ser2009Pogorelov, ser2013Pogorelov, ser2017Pogorelov} that have also explored this topic.


The work structure is as follows: in Section~\ref{M} we describe the model, and numerical approach, and also formulate the problem in dimensionless form, Section~\ref{R} presents the results and discussions, Section~\ref{C} summarizes the results and discusses plans for future work, \ref{ap1} offers additional distributions of hydrogen atoms for a more complete description of the obtained results, \ref{ap2} presents a two-dimensional picture of the flow and position of the discontinuity surfaces.

\section{Model}

\label{M}

In this work, it is assumed that the interstellar medium consists of two components: an ionized component and a neutral component consisting of hydrogen atoms. The ionized component is considered to be a mixture of protons and electrons with the assumption of quasi-neutrality and the equation of state: $p_p = 2 n_p k T_p$. The motion of such a mixture is described by a system of Euler equations for a monoatomic non-thermal-conducting perfect gas with constant heat capacities ($\gamma = 5/3$):

\begin{align}
\begin{cases}
\dfrac{\partial \rho_p}{\partial t} + \textbf{div}( \rho_p{\bf V}_p) = 0,\\[3mm]
\dfrac{\partial (\rho_p {\bf{V}}_p)}{\partial t} + \textbf{div}( \rho_p{\bf V}_p {\bf V}_p + p_p\hat{\bf I}) = \bf Q_2,\\[3mm]
\dfrac{\partial E_p}{\partial t} + \textbf{div}( (E_p+p_p) {\bf V}_p) = Q_3,\\[3mm]
\end{cases}
\label{sys1}
\end{align}

where
$
E_p = \frac{p_p}{\gamma-1}+ \frac{\rho_p V_p^2}{2}
$ - is the energy density; $\rho_p,\ p_p,\  {\bf{V}}_p$ - the plasma density, pressure and velocity, respectively.

The interaction of the neutral component with the plasma is taken into account in the right parts of the equations of motion and energy. 
For the interstellar hydrogen, we assume kinetic description. The Boltzmann kinetic equation for the velocity distribution function $f_\mathrm{H}$ of hydrogen atoms is solved:
\begin{align}
	&\frac{\partial f_\mathrm{H}}{\partial t} + {\bf V_\mathrm{H}}\cdot \dfrac{\partial f_\mathrm{H}}{\partial \bf r} + \dfrac{\bf F}{m_\mathrm{H}} \cdot \dfrac{\partial f_\mathrm{H}}{\partial {\bf V_\mathrm{H}}} =  \notag \\
	 &= -f_\mathrm{H}\ n_p \int u\ \sigma_{ex}^{\mathrm{HP}}(u)\ f_p({\bf V}_p)\ d{\bf V}_p\ + \label{kma} \\
	&+ f_p({\bf V_\mathrm{H}})\ n_p \int |{\bf V}_\mathrm{H}^* - {\bf V}_\mathrm{H}| \sigma_{ex}^{\mathrm{HP}}(|{\bf V}_\mathrm{H}^* - {\bf V}_\mathrm{H}|) f_\mathrm{H}({\bf V}_\mathrm{H}^*) d{\bf V}_\mathrm{H}^* \notag,
\end{align}
here and further ${\bf u} = {\bf V}_\mathrm{H} - {\bf V}_p$, $u = |{\bf u}|$. ${\bf V}_p,\ {\bf V}_\mathrm{H}$ - individual velocities of protons and hydrogen, respectively. {$\bf F$} - the total force of gravity and radiation pressure of the star.
$\sigma_{ex}^{\mathrm{HP}}(u) = (2.2835 \cdot 10^{-7} - 1.062 \cdot 10^{-8} \mathrm{ln}(u))^2$ - (cm$^2$) the charge exchange cross section ($u$ is in cm/s, see \cite{Lindsay2005}).

Here and below, for simplicity, it is assumed that the influence of the force {$\bf F$} on the global flow pattern is insignificant. This means that hydrogen atoms fly along rectilinear trajectories between  charge exchanges with protons. In fact, in the case of the heliosphere, the deviation of trajectories from straight lines is significant only at distances close to the star (several astronomical units, see \cite{izmodenov_alexashov2015}), which can be neglected in the global problem (with a characteristic scale of several hundred astronomical units).

The velocity distribution function of plasma protons $f_p$ is assumed to be locally Maxwellian:
\begin{align}
f_p({\bf V}_p) & = (\sqrt{\pi} c_p)^{-3} \mathrm{exp}\left( -\dfrac{ ({\bf V}_p - {\bf U}_p)^2}{c_p^2}\right),\\
c_p & = \sqrt{\dfrac{2k_B T_p}{m_p}}, \nonumber
\end{align}
where ${\bf U}_p,\ T_p$ - mass velocity and plasma temperature, respectively; $k_B$ - is the Boltzmann's constant.

The expressions for the sources of momentum and energy in plasma (see System~\ref{sys1}) can be written as follows:
\begin{align}
\begin{cases}
{\bf Q}_2 & = n_\mathrm{H}\cdot \rho_p\cdot \iint u \cdot \sigma_{ex}^{\mathrm{HP}}(u) \cdot \ {\bf u} \ \cdot f_\mathrm{H}({\bf V}_\mathrm{H})\\
&\cdot f_p({\bf V}_p) d{\bf V}_\mathrm{H} d{\bf V}_p,\\
Q_3 & = n_\mathrm{H}\cdot \rho_p\cdot \iint u\cdot \sigma_{ex}^{\mathrm{HP}}(u) \cdot \dfrac{{\bf V}_\mathrm{H}^2 - {\bf V}_p^2}{2} \cdot f_\mathrm{H}({\bf V}_\mathrm{H}) \\
&\cdot  f_p({\bf V}_p)d{\bf V}_\mathrm{H} d{\bf V}_p. 
\end{cases}
\label{syssource}
\end{align}

In this work, the influence of the interstellar magnetic field is neglected to make the parametric study feasible.

\subsubsection{Boundary conditions}

Let's describe the boundary conditions for both plasma and neutral component. We connect the origin of the coordinate system with the star and the X-axis is chosen toward the incoming flow of the interstellar medium. The star is considered to be the hypersonic source flow of the fully ionized hydrogen plasma (the Mach number $M \gg 1$) with a given mass lose rate $\dot{M}_\star = 4\pi\rho V_0 R^2$ and the terminal velocity $V_0$. The interstellar medium is considered as a parallel flow of similar plasma with density $\rho_{p,\infty}$, velocity $V_{\infty}$, and pressure $p_{p,\infty}$.
 
 The kinetic equation for the neutral component (atomic hydrogen) is hyperbolic. Therefore, the velocity distribution function should be set at the boundaries only for incoming characteristics.
 We suppose that the velocity distribution function of Maxwellian:
 \begin{align}
& f_{\mathrm{H}, \infty}({\bf V}_\mathrm{H})  = (\sqrt{\pi} c_\infty)^{-3} \mathrm{exp}\left( -\dfrac{ ({\bf V}_\mathrm{H} - {\bf V}_\infty)^2}{c_\infty^2}\right),\\
& c_\infty  = \sqrt{\dfrac{2k_B T_\infty}{m_p}}, \nonumber
\end{align}
where $T_\infty,\ {\bf V}_\infty$ - temperature and velocity of the unperturbed local interstellar medium (here it is assumed that the temperature and velocity of hydrogen at infinity are respectively equal to the temperature and velocity of the plasma). The hydrogen number density is $n_{\mathrm{H},\infty}$.

 \subsection{Dimensionless formulation of the problem}
\label{Dim}
 The formulated above problem depends on seven independent parameters: $\dot{M}_\star,\ V_0,\ \rho_{p,\infty},\ c_{\infty},\ V_{\infty},\  n_{\mathrm{H}, \infty},\ \sigma_{ex, \infty}^{\mathrm{HP}} = \sigma_{ex}^{\mathrm{HP}}(c_\infty)$. The latter parameter is the charge exchange cross-section corresponding to the thermal velocity $c_\infty$. The value of this parameter is known from the expression for the cross-section given above (see Section~\ref{M}).

 We can reformulate the problem in dimensionless form by reducing the number of parameters to four. Let us relate all distances to $L = \sqrt{\dfrac{0.78\cdot \dot{M}_\star V_0}{4 \pi \rho_{\mathrm{p},\infty} c^2_{\infty}}}$, all velocities to the thermal velocity $c_\infty$, the plasma density to $\rho_{p,\infty}$, the atom number density to $n_{\mathrm{H}, \infty}$. The constant 0.78 in the definition of $L$ is a numerical result used for convenience, as in this case, the dimensionless distance to the bow shock in the purely gas-dynamic case (for Mach number is 1.97) is approximately equal to $1$ ($L \sim L_{BS}$).

This choice of characteristic length is based on the analogy with the analytically derived distance to the discontinuity surface in the thin layer approximation when the interstellar medium Mach number is much greater than 1, as described in the works of \cite{Baranov1970} and \cite{1971Baranov}: $L_{0} = \sqrt{\dfrac{\dot{M}_\star V_0}{4 \pi \rho_{\mathrm{p},\infty} V^2_{\infty}}}$. However, it is well known now that the shock layer is not thin, for the value of 1.97 of Mach number (for the heliosphere), so this formula does not provide an exact distance for any of the discontinuity surfaces.




The System~\ref{sys1} in dimensionless form is as follows:
\begin{align}
\begin{cases}
\dfrac{\partial \hat{\rho}_p}{\partial \hat{t}} + \textbf{div}( \hat{\rho}_p \hat{{\bf V}}_p) = 0,\\[3mm]
\dfrac{\partial (\hat{\rho}_p \hat{{\bf V}}_p)}{\partial \hat{t}} + \textbf{div}( \hat{\rho}_p \hat{{\bf V}}_p \hat{{\bf V}}_p + \hat{p}_p\hat{\bf I}) = \dfrac{\eta}{\mathrm{Kn}_\infty} \cdot \hat{\bf Q}_2,\\[3mm]
\dfrac{\partial \hat{E}_p}{\partial \hat{t}} + \textbf{div}( (\hat{E}_p+\hat{p}_p) \hat{{\bf V}}_p) = \dfrac{\eta}{\mathrm{Kn}_\infty} \cdot \hat{Q}_3,\\[3mm]
\end{cases}
\label{sys2}
\end{align}

Kinetic Equation~\ref{kma} ($\bf F = 0$) is as follows:
\begin{align}
	&\frac{\partial \hat{f}_\mathrm{H}}{\partial \hat{t}} + {\hat{\bf V}_\mathrm{H}}\cdot \dfrac{\partial \hat{f}_\mathrm{H}}{\partial \hat{\bf r}} =  \notag \\
	 &= -\dfrac{\hat{f}_\mathrm{H}}{\mathrm{Kn}_\infty} \int |\hat{\bf V}_\mathrm{H} - \hat{\bf V}_p| \hat{\sigma}_{ex}^{\mathrm{HP}}(|\hat{\bf V}_\mathrm{H} - \hat{\bf V}_p|) \hat{f}_p(\hat{\bf r},\hat{\bf V}_p) d\hat{\bf V}_p\ + \label{kma2} \\
	&+ \dfrac{\hat{f}_p(\hat{\bf r}, {\hat{\bf V}_\mathrm{H}})}{\mathrm{Kn}_\infty} \int |\hat{\bf V}_\mathrm{H}^* - \hat{\bf V}_\mathrm{H}| \hat{\sigma}_{ex}^{\mathrm{HP}}(|\hat{\bf V}_\mathrm{H}^* - \hat{\bf V}_\mathrm{H}|) \hat{f}_\mathrm{H}(\hat{\bf r},\hat{\bf V}_\mathrm{H}^*) d\hat{\bf V}_\mathrm{H}^* \notag
\end{align}

Dimensionless boundary conditions will now be written as follows:

$$\hat{V}_0 = \chi,\ \hat{\dot{M}}_\star = 4\pi/\chi,\  \hat{\rho}_{p,\infty} = 1,\ \hat{c}_\infty = 1,$$ 
$$ \hat{V}_{\infty} = M_{\infty} \sqrt{\gamma},\ \hat{n}_{\mathrm{H}, \infty} = \eta.$$

The dimensionless charge exchange cross section is: 
\begin{align}
&\hat{\sigma}_{\mathrm{ex}}(\hat{U}) = \left(1 - \hat{a}_2 \cdot \mathrm{ln}(\hat{U}) \right)^2, \\
& \hat{a}_2 = \dfrac{1.062\cdot 10^{-8}}{\sqrt{\sigma_{\mathrm{ex}}^\mathrm{HP}(c_\infty)}}. \nonumber
\end{align}
As a result, four dimensionless parameters of the problem are obtained:
$$\chi,\ \eta,\ M_\infty,\  \mathrm{Kn}_\infty.$$
Descriptions of these parameters are following: \\
(1) $\chi = \dfrac{V_0}{C_\infty}$ is the ratio of the terminal velocity of the star to the thermal velocity of the incoming flow. This parameter appeared in the dimensionless formulation of the inner boundary condition. This parameter does not affect the geometric pattern of the flow in a purely gas-dynamic case (see, for example, \cite{Korolkov2020}). However, decreasing this parameter leads to an increase in plasma number density inside the astrosphere and an effective increase in the charge exchange frequency, which strongly affects the distribution of hydrogen atoms, and hence the global structure as a whole.\\
(2) $\eta = \dfrac{n_{\mathrm{H}, \infty}}{n_{p, \infty}}$ is the ratio of hydrogen number density to proton number density in the incoming flow. This parameter linearly affects the magnitude of momentum and energy sources (see System \ref{sys2}), but the Equation~\ref{kma2} does not depend on this parameter. 
\\
(3) $M_\infty$ is the Mach number of the incoming flow. It determines the velocity of the interstellar medium and, consequently, astropause position and shape. In \cite{Korolkov2020}, the dependence on this parameter in a purely gas-dynamic case has been studied. The influence on the structure of the astrosphere in the presence of neutral atoms has not yet been investigated. \\
(4) $\mathrm{Kn}_\infty = \dfrac{l_{ex, \infty}}{L}$ is the Knudsen number, the ratio of the mean free path of an atom to the characteristic size of the problem.  The mean free path is calculated as follows:
\begin{align}
l_{ex, \infty} = \dfrac{1}{n_{p, \infty}\  \sigma^{\mathrm{HP}}_{\mathrm{ex}}(c_\infty)}.
\end{align}
From System~\ref{sys2} and Equation~\ref{kma2}, it is obvious that the lower the Knudsen number, the greater the right part sources magnitude, and the more significant the influence of the charge exchange on the flow.

To perform a complete parametric study is unfeasible because even a single calculation is time-consuming. This work focuses on the dependence of the solution on the Knudsen number. Physically, variation of this dimensional parameter (with simultaneous keeping of all other parameters) means variation in the mass loss of the star $\dot{M}_{\star}$ (by changing the number density of the stellar wind).


A fairly wide range of values of $\mathrm{Kn}_\infty$ from $10^{-4}$ to $10^{2}$ is presented in this paper. The other three parameters are fixed and their values are chosen to be close to those of the heliosphere: $\chi = 36.2, \, \eta = 3,\, M_\infty = 1.97$. For heliospheric parameters, the $\mathrm{Kn}_\infty = 0.43$ ($n_{p, \infty} = 0.073\ \mathrm{cm}^{-3},\ V_{\infty} = 26.4\ \cdot 10^5\ \mathrm{cm/s},\ T_\infty = 6500\ \mathrm{K},\ n_{p, E} = 7.3\ \mathrm{cm}^{-3}, \ V_E = 375\ \cdot 10^5\ \mathrm{cm/s}$), L = $319$ AU.

\subsection{Numerical approach}

This subsection briefly describes the numerical methods to solve the problem. Our approach consists of two fundamentally different parts that are solved sequentially within the framework of the global iterations. The first part involves the numerical solution of the system of Equations~(\ref{sys1}), assuming that the momentum and energy sources on the right-hand sides of the equations can be calculated by using the numerical solution of the kinetic equation obtained in the second part (for the first iteration the sources are assumed to be zero). Then, using the plasma distributions obtained in the first part, the kinetic Equation~(\ref{kma}) is solved by the Monte Carlo method, and new momentum and energy sources are obtained. The iterative process is repeated until the plasma and atom distribution are fully established. Some details of the Monte-Carlo method can be found in \cite{Sobol}, and the Moscow model algorithm can be found in \cite{Malama1991}.

\begin{figure*}
	\includegraphics[width=\hsize]{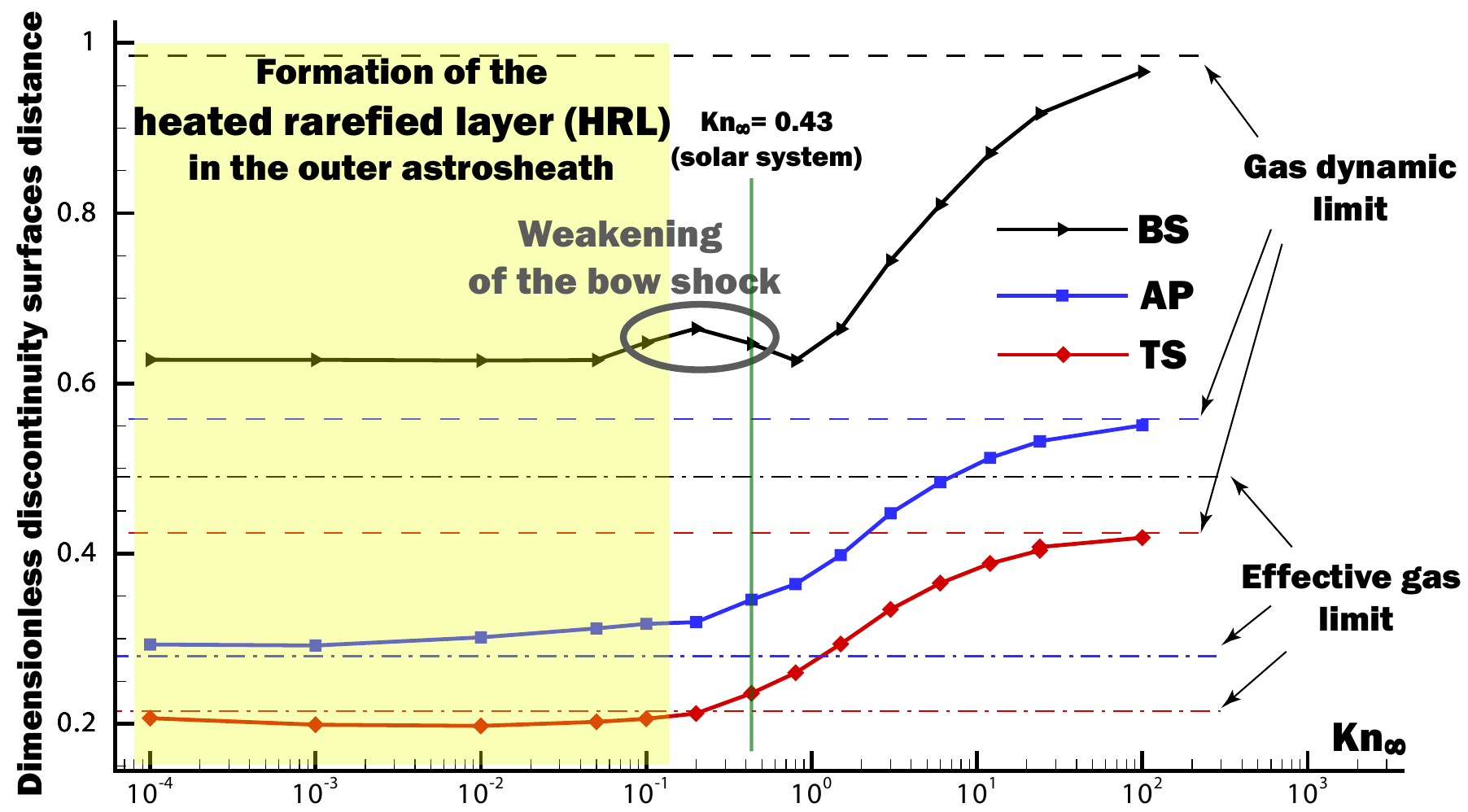}
	\caption{Dimensionless distance from the star to the discontinuity surface on the X-axis (upwind) for various values of the Knudsen number. The internal shock wave (TS) is marked in red, the astropause (AP) - in blue, and the external shock wave (BS) - in black. Horizontal dotted lines show the positions of the surfaces for the gas-dynamic solution}
	\label{fig0}
\end{figure*}

To solve the System~\ref{sys1}, the finite volume method was chosen, described in detail in \cite{godunov1976}. The system is solved by the time-relaxation method. The Riemann problem at the boundaries of numerical cells is solved using both the approximate HLLC solver (see \cite{Miyoshi}) and the exact solver by \cite{godunov1976}. The minmod limiter (see \cite{1990Hirsch}, Formula 21.3.23a) is used to obtained linear interpolation of the gas parameters within a cell.
The source terms (${\bf Q}_2,\ Q_3$) are calculated in Monte-Carlo code for $\mathrm{Kn}_\infty > 0.05$.
For $\mathrm{Kn}_\infty <= 0.05$, the Monte Carlo method may provide insufficient statistics for the sources, so we calculate the average values of the number density, velocity, and temperature of atoms in the computational cells by the Monte Carlo method and then, employ the formulas for the source terms obtained in the \cite{McNutt}.


All calculations have been performed using a two-dimensional computational grid that captures the main discontinuity surfaces: the termination shock (TS), the tangential discontinuity (i.e. astropause, AP), and the bow shock (BS). The grid consists of approximately 15000 cells and its head region resembles a sphere. An example of such a grid is shown in Figure~2, A of \cite{izmodenov_alexashov2015}. In the inner shock layer between the TS and AP, the number of cells along the radial direction is 30, while in the outer shock layer and the super-sonic stellar wind region, it is 40. In the interstellar medium, it is 35, with a finer resolution towards BS. The number of cells in the angular direction (from 0 to $\pi$) is 100. Additional test calculations were also performed on a finer (2 times in each direction) grid to validate the results.

The kinetic Equation~(\ref{kma}) was solved also using the same grid. However, since atoms move in three-dimensional space, the cells for them are formed by rotating of two-dimensional cells around the axis of symmetry. In this way, the kinetic sources were averaged over azimuthal direction. The spatial extent of the grid in dimensionless variables is 5.5 in the upwind direction, 4 in the downwind direction, and 5.5 in the perpendicular direction. It should be noted that the statistic of the Monte Carlo method linearly depends on the volume of the cells, so it is necessary to find a balance in the grid resolution between gas dynamics and the kinetic equation.

At the inlet boundaries, so-called rigid boundary conditions are chosen for the plasma parameters, i.e. all parameters (density, speed, and pressure) are fixed. So-called soft boundary conditions were used at outlet boundaries (i.e. the derivatives of all parameters are assumed to be zero). It has been verified in numerical tests that the boundary conditions do not impose additional disturbances on the flow.

\subsection{Limiting solutions}
\label{Lim}

It is quite natural to expect that the problem formulated above has two limiting cases: $\mathrm{Kn}_\infty \rightarrow \infty$ and $\mathrm{Kn}_{\infty} \rightarrow 0$. In these cases, the solution to the problem becomes simpler. In the case where $\mathrm {Kn}_{\infty}\rightarrow\infty$, atoms do not charge exchange with the charged component. The sources of momentum ($\bf{Q}_2$) and energy (Q3) for the plasma become zero (see System~\ref{sys2}). As a result, we obtain a simple gas dynamic solution for protons. We call this case the plasma gas dynamic limit.

In the case of $\mathrm{Kn}_\infty~\rightarrow~0$, the mean-free path $l$ is much much smaller than the characteristic size $L$, so in any small volume, the plasma and neutrals exchange their momenta and energy very effectively.
Therefore, both velocities and temperatures of the components are identical in the entire heliosphere.
In this case so-called effective gas approximation may be employed. In this approximation a solve system of Equations~(\ref{sys2}) for the mixture of plasma and atoms with the following boundary conditions:
$\rho_\infty = \rho_{p, \infty} + \rho_{\mathrm{H},\infty}$ ($\rho_\mathrm{H} = m_p \cdot n_\mathrm{H}$), $p_\infty = (2 n_{p, \infty} + n_{\mathrm{H}, \infty}) k_B T_\infty$. The parameters for the stellar wind do not change. We call this case the effective gas limit. In the final solution, the ratio of plasma and atom number densities is maintained, so it is easy to separate protons from hydrogen: $\rho_p = \rho \cdot \rho_{p, \infty} / \rho_\infty$.

\subsection{Local Knudsen number}
\label{Local_Knudson}

Along with the Knudsen number $\mathrm{Kn}_\infty$ defined in Subsection~\ref{Dim}, we can introduce local Knudsen numbers (or the local mean free paths of an atom):
\begin{align}
\mathrm{Kn} = \dfrac{l_{ex}}{L}, \ \ l_{ex} = \dfrac{V_\mathrm{H}}{\nu_{ex}}, \nonumber \\ 
\nu_{ex} = n_p\int u \cdot \sigma^{\mathrm{HP}}_{\mathrm{ex}}(u) \cdot f_p ({\bf r, V}_p) d{\bf V}_p,
\label{LocKn}
\end{align}
where $\nu_{ex}$ is the charge-exchange rate, $V_H$ is the hydrogen atom velocity. The local Knudsen number depends on the velocity of the considered atom and on the local properties of the plasma. Note that the dependence of the charge exchange rate on the local plasma number density is linear.

Let us consider the local Knudsen number in the heliosphere (or similar astrosphere) for an interstellar neutral atom moving with the speed of the interstellar medium. In the undisturbed LISM the local Knudsen number is equal to $\mathrm{Kn}_\infty$.
Due to heating and increased plasma number density in the outer shock layer after passing through the BS, the mean free path of the interstellar atom decreases by approximately four times.  When the atom enters the inner shock layer,  the local Knudsen number immediately increases by approximately ten times due to low plasma number density. It then enters the region of the supersonic stellar wind, the mean free path increases approximately two times, since the plasma density in front of the TS is smaller than behind it. As the atom approaches the star, solar wind number density increases proportionally to the square of the distance to the star, reaching high values. The mean free path at 1 AU decreases by 10000 times compared to the distant heliosphere. In fact, the atoms do not reach such close distances but charge exchange earlier, forming an atom-free zone.


It is worthwhile to note that the local mean free paths (and, therefore, local Knudsen numbers) for newly created secondary neutrals can be significantly different from  those described above, because of their different velocities. For example, the mean free path of an atom of a third population (born in the outer shock layer, see Section \ref{atom_par} for more information about populations) can be five times shorter in the inner shock layer than for interstellar atoms in this region. On the other hand, atoms of the first population (supersonic neutral wind) have eight times greater mean free paths for the inner shock layer compared to interstellar atoms in this region. These considerations highlight the complexity of the problem and the importance of the kinetic description for taking into account all populations of atoms and their influence on a particular flow region.

\section{Results and discussion}
\label{R}

\begin{figure*}
	\includegraphics[width=\hsize]{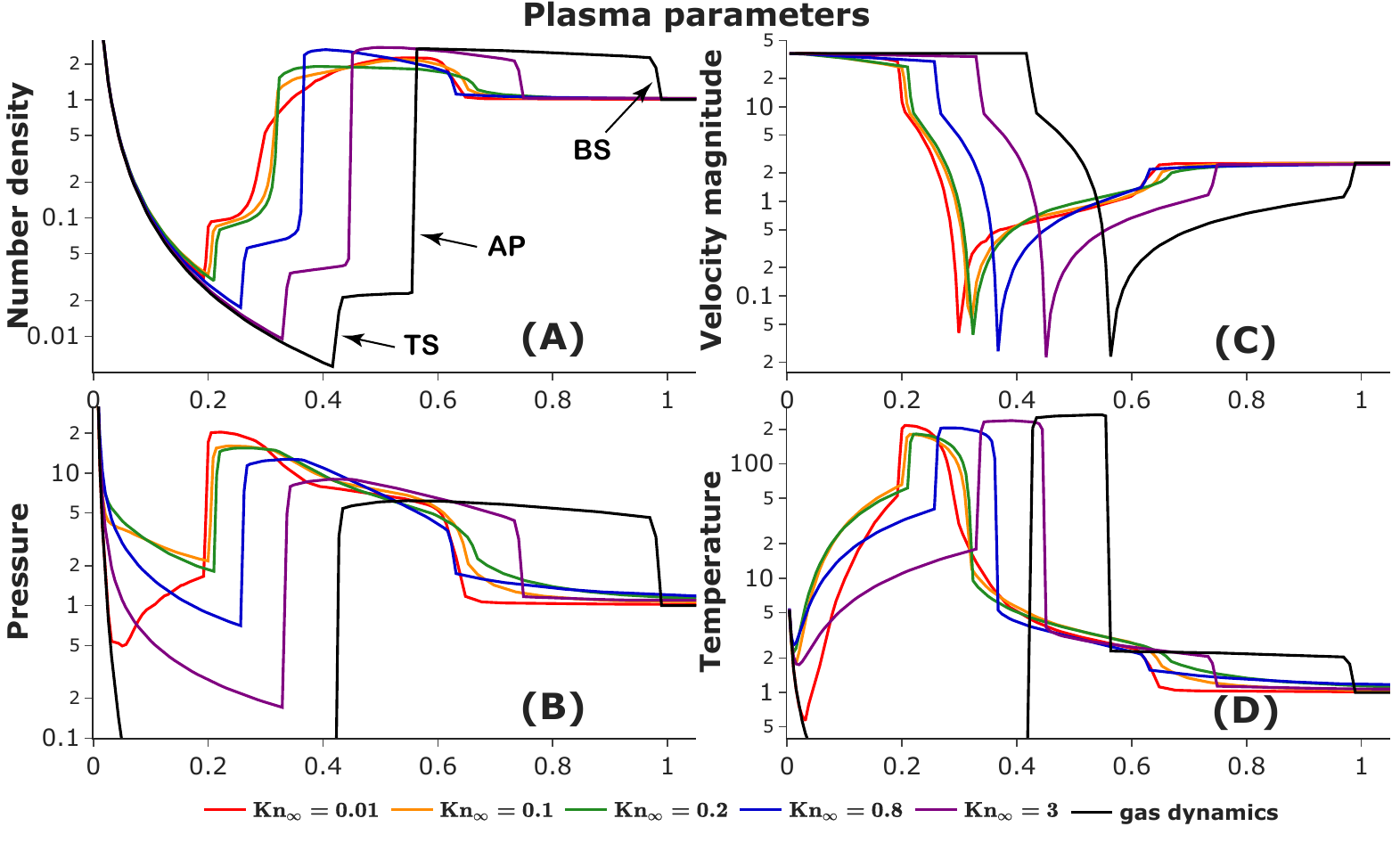}
	\caption{Dimensionless distributions of number density (A), pressure (B), velocity magnitude (C), and temperature (D) on the X-axis (upwind) for various values of the Knudsen number}
	\label{fig1}
\end{figure*}

\subsection{Plasma parameters}

In this section, we present results obtained in the frame of our model. Figure~\ref{fig0} is a kind of general summary of all results. It presents the dimensionless distances from star to the three discontinuity surfaces (TS, AP, and BS) along the X-axis (the upwind direction) for various values of the Knudsen number. The dots represent the results of the calculations. The value of $\mathrm{Kn}_\infty \approx 0.43$ corresponds to the case of the heliosphere. It is marked by the vertical green line.  

Horizontal dash and dash-dot lines correspond to the plasma gas-dynamic and effective gas limits, respectively. Based on our results, we were able to numerically determine the positions of the discontinuity surfaces in the pure gas-dynamic case for $M_\infty = 1.97$ and derive the following formula:
\begin{align}
L_S = k \cdot \sqrt{\dfrac{0.78 \cdot \dot{M}_\star V_0}{4 \pi \rho_{\mathrm{p},\infty} c^2_{\infty}}},
\label{LLL}
\end{align}
where $k = 0.42, 0.56, 0.98$ for $L_S = L_{\textrm{TS}},\ L_{\textrm{AP}}, \textrm{and}\ L_{\textrm{BS}}$, respectively.
In principle, it is expected that the positions of the surfaces will tend to these limits for large and small values of the Knudsen numbers. As seen in Figure~\ref{fig0}, the numerical solution reaches the upper limit at  $\mathrm{Kn}_\infty = 10^2$. This shows that for larger values of Knudsen numbers, the influence of charge exchange becomes negligible and the plasma gas dynamic limit can be used.

For the small values of the Knudsen number ($\mathrm{Kn}_\infty \le 0.2$), the positions of the TS and AP approach those obtained in the effective gas limit. However, the position of the BS does not follow this trend. Even for extremely small values of the Knudsen number  ($\mathrm{Kn}_\infty \sim 10^{-4}$) the numerical solution does not converge to the effective gas limit, due to the formation of a specific flow structure in the outer shock layer. We call this structure the heated rarefied layer and describe it in detail in Subsection~\ref{PDL}. Additionally, the flow pattern at extremely low values of Knudsen numbers is discussed in \ref{ap1}.


Note, that our model cannot accurately describe the solution for values of $\mathrm{Kn}_\infty < 10^{-4}$. The Monte Carlo method has computational limitations and cannot handle mean free paths near zero.


The dimensionless distances to the shocks and the astropause increase monotonically with increasing Knudsen numbers. The exception is the bow shock distance within the range of $10^{-1} \leq \mathrm{Kn}_\infty \leq 0.43$, which is marked by a gray ellipse in Fig.~\ref{fig0}. It is interesting to note that for these values of Knudsen, the intensity of the velocity and density jumps at the shock is extremely low. This indicates that there is weakening of the bow shock, which will be discussed further in Subsection \ref{shockless}.


Figure~\ref{fig1} presents the plasma number density (panel A), pressure (panel B), velocity magnitude (panel C), and temperature (panel D) along the X-axis (upwind) for various values of the Knudsen number. The black curve corresponds to the plasma-gas limit. For this limit, the positions of TS, AP, and BS are marked in the figure (see panel A). For the plasma-gas limit, the maxima of the number densities (panel A, black curve) are at the astropause for both outer (i.e. between BS and AP) and inner (between TS and AP) shock layers.
However, as the Knudsen number decreases, the maximum moves in the outer shock layer towards the bow shock. For $\mathrm{Kn}_\infty = 0.1$ (orange curve) and below, the number density decreases towards the astropause in almost the entire shock layer, reaching its minimum at the astropause. We called this effect the heated rarefied layer of plasma (HRL). We observed HRL in the range of Knudson numbers: $10^{-4} \leq \mathrm{Kn}_\infty \leq 0.1$ (the range marked in yellow in Figure~\ref{fig0}). The physical reasons for HRL are discussed in detail in Subsection~\ref{PDL} (also see schematic diagram in Figure~\ref{shem1}, B).  The ratio of maximum and minimum number density in the outer shock layer is quite strong, approximately $3.8$ for $\mathrm{Kn}_\infty  = 0.01$ (panel A, red curve). On the contrary, in the inner shock layer, the number density increases towards the astropause more strongly for smaller Knudsen numbers. 


Similar to density, the pressure has a maximum at the astropause in plasma-gas limit (see Figure~\ref{fig1}, B, black curve). The astropause is not visible in the pressure profile, since the pressures from both sides are equal. It is located at $X \approx 0.56$. 
In the outer shock layer, the pressure maximum remains at the astropause for all Knudsen numbers. In the inner shock layer, the maximum pressure moves out of the astropause and approaches the termination shock at $\mathrm{Kn}_\infty \leq 0.2$ (green, orange, and red curves). It is worth noting that for $\mathrm{Kn}_\infty  = 0.01$ the pressure increases with the distance in the supersonic stellar wind region starting at distances $\approx 0.05$ (red curve, panel B).

The velocity profiles (Figure~\ref{fig1}, C) demonstrate the deceleration of the supersonic stellar wind before the TS. This effect has long been known for the solar wind in the literature  (see, e.g., \cite{2006Izmodenov}, Figure 4.3). The reason for the deceleration is an effective loss of momentum as a result of charge exchange. Interestingly, the greatest deceleration occurs at $\mathrm{Kn}_\infty  \sim 0.1 - 0.2$ (orange and green curves, panel C). At smaller values of $\mathrm{Kn}_\infty  = 0.01$, the atoms influence the plasma velocity only in the immediate vicinity of the termination shock, causing weaker deceleration.

The temperature profiles (Figure~\ref{fig1}, D) are consistent with the density and pressure and with the equation of state of the plasma ($p_p = 2 n_p k_B T_p$). In a supersonic stellar wind, temperature increases strongly towards the termination shock. This increase is due to charge exchange, as (in our approach) newly injected protons (called pickup protons) have large thermal velocities in the stellar wind rest frame.
This effect is well-known for the heliosphere (see, e.g., \cite{1994Gazis, 1995Lazarus}). The lower the Knudsen number, the stronger the heating effect. It is also interesting to note the peculiarity at $\mathrm{Kn}_\infty  = 0.01$ (panel D, red curve) that the plasma begins to heat up further from the star than at, for example, with $\mathrm{Kn}_\infty  = 0.1$ (panel E, orange curve). Nevertheless, the temperatures near the terminal shock is almost the same. This is because for small Knudsen numbers coupling between protons and atoms is stronger and atoms do not penetrate closer to the star. As a result, the atom-free zone appears (see details of the atomic distribution in Subsection~\ref{atom_par}).
It is worthwhile to note that the strong increase in the temperature of the solar wind appeared in the so-called one-fluid approach with all charge components (electrons, original solar protons, and pick-up protons) being treated as a single fluid. 
For the heliosphere, however, the pick-up protons do not completely assimilate, so their temperature differs from that of the solar winds protons (see, for example, \cite{2006Malama}, \cite{2022Korolkov}).

In the outer shock layer, the temperature increases towards the astropause. This occurs more intensely at lower Knudsen numbers. The plasma in the outer astrosheath is heated by the action of ENA, which causes the appearance heated rarefied layer of plasma (see subsection~\ref{PDL}).


\subsection{Atom parameters}
\label{atom_par}

\begin{figure*}[h]
	\includegraphics[width=\hsize]{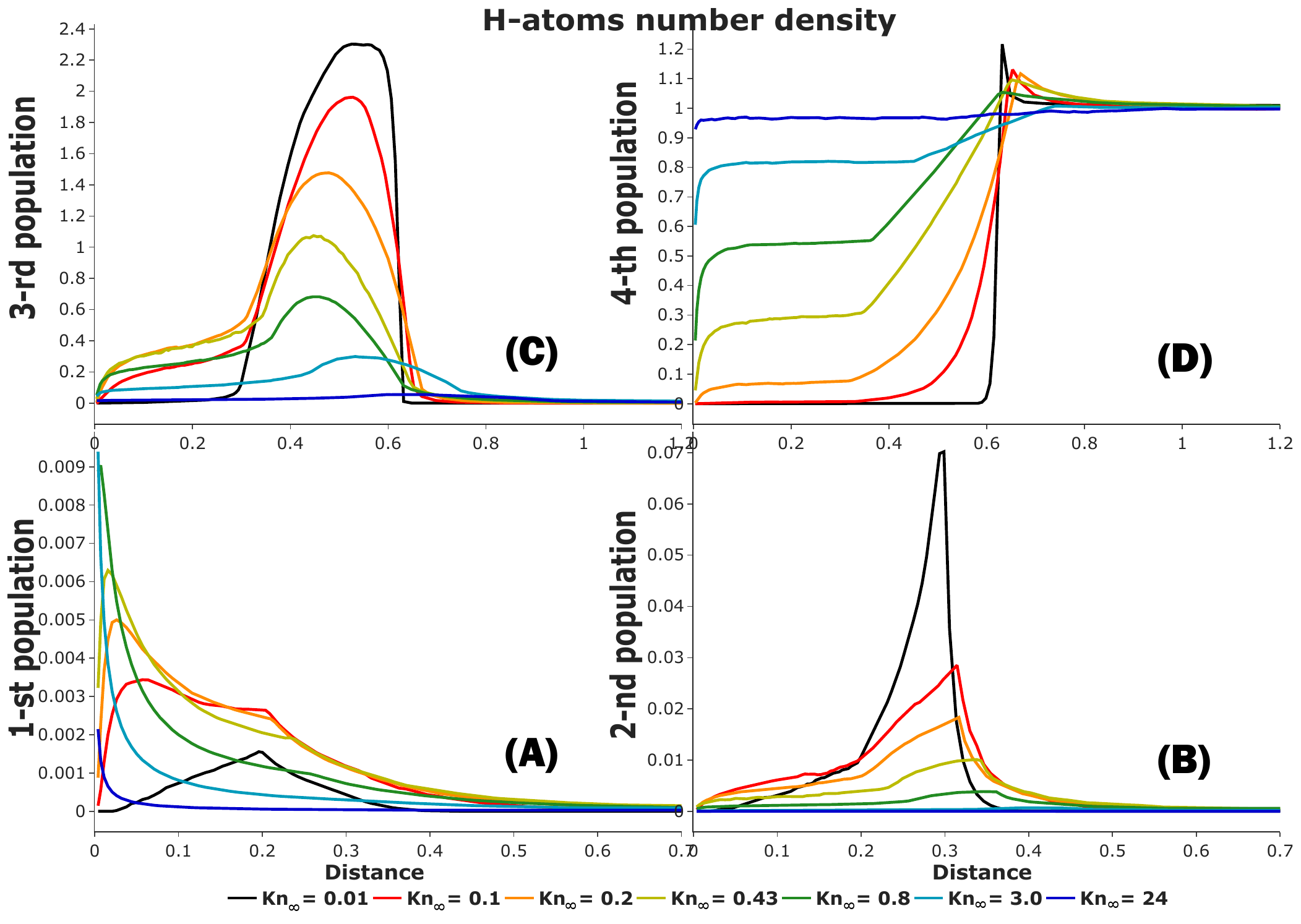}
	\caption{Number density of each population of hydrogen atoms for different values of the Knudson number on the X-axis (upwind). Values are dimensionless for hydrogen concentration in LISM}
	\label{fig2}
\end{figure*}

In this section, we present the distributions of atom number densities in astrospheres. To better understand the charge exchange process, we divide all atoms into four populations according to \cite{2000Izmodenov}. Population 4 represents interstellar hydrogen or atoms born in the supersonic interstellar medium. Population 3 is atoms born in the outer astrosheath. Population 2 is atoms born in the inner astrosheath (ENA). Population 1 is atoms born in the supersonic stellar wind (aka neutral stellar wind).

Figure~\ref{fig2} shows the distributions of the number density of each Population (1 - 4) of atoms for different values of the Knudsen number on the X-axis (upwind). First of all, let us pay attention to the distribution of atoms in the heliosphere case ($\mathrm{Kn}_\infty = 0.43$, yellow curves). In this case, the bow shock is located at the distance $X \approx 0.65$. The number density of Population 4  (panel D) remains almost constant over a large distance ($X > 0.9$), and increases slightly before the bow shock in the region $0.65 \leq X \leq 0.9$. This slight increase in number density in a small region, whose length is on the order of the atom mean free path, is primarily due to the atoms of Population 3 that fly outward from the outer shock layer. Strictly speaking, ENAs also penetrate this region, but, as will be shown below, their number density is extremely low at these distances, so their contribution to the number density increase is negligible, what cannot be said about their contribution to the source of momentum and energy for the plasma, due to their high energies. \cite{1982Gruntman} assessed the effect of ENA on plasma in the supersonic interstellar medium and showed the possibility of shockless transition (we discuss this in Subsection~\ref{shockless}). Note that the increase in Population 4's number density can be considered as a precursor for the hydrogen wall. We will discuss this next. 

After passing through the bow shock, the fourth population begins to effectively decrease in the outer shock layer ($0.35 \leq X \leq 0.65$, panel D, yellow curve) due to charge exchange. As a result, atoms of Population 3 are born in this region (panel C). These atoms form the hydrogen wall, predicted by \cite{BaranovLebedev1991} and detected in the direction of $\alpha$ Cen \citep{Linsky1996ApJ...463..254L} on the Hubble Space Telescope.

After crossing the astropause the atoms of Populations 3 and 4 enter the inner shock layer ($0.23 \leq X \leq 0.35$). In this case, atoms of Population 2 are originated (Figure~\ref{fig2}, B). We also note that charge exchange in the outer shock layer occurs more efficiently than in the inner one (this can be seen, for example, from the slope of the yellow curve in panel D). It means that the local Knudsen number (see Subsection~\ref{Local_Knudson}) in the outer shock layer is significantly higher than one in the inner layer.  This is due to the low number density of plasma in the inner astrosheath.

Atoms of Population 1 are born in the supersonic solar wind ($X \leq 0.23$, panel A). Their maximum number density in the heliospheric case is achieved at a distance $X \approx 0.03$. With decreasing distance to the star, the plasma number density increases as the square of the distance ($n_\mathrm{p} \sim 1/r^2$), as a result of which the mean free path of atoms similarly decreases. This leads to the fact that at a certain distance from the star, all atoms undergo charge exchange, and newborn atoms (neutral solar wind) spread out from the star. Thus, near the star, there is a region without atoms, which we call the atom-free zone.

Now, we will explore how the distribution of the populations depends on the Knudsen number. Firstly, the number density of atoms in Population 4 (panel D) inside the astrosphere becomes smaller as the Knudsen number decreases. Their number density also decreases in the outer astrosheath. However, the increase in number density before the bow shock (a precursor to the hydrogen wall, $X \approx 0.62$) becomes significantly larger. For example, for $\mathrm{Kn}_\infty = 0.01$, the increase in density is more than 20$\%$ of the interstellar value. The width of this region becomes narrower, which is quite expected since the width is on the order of the mean free path of an atom, which decreases with decreasing Knudsen number. Note that for the case of $\mathrm{Kn}_\infty = 0.01$, almost all of Population 4 undergoes charge exchange after passing through the bow shock.

The height of the hydrogen wall (panel C) increases with decreasing Knudsen number, which was predicted by the precursor of the hydrogen wall. The number density of atoms from 3rd population that penetrated the astrosphere (at $X < 0.35$) also decreases. The number density of atoms of Population 2 (panel B) becomes larger in the inner astrosheath. In the supersonic stellar wind for $\mathrm{Kn}_\infty \geq 0.1$, the atom number density of Population 2 also increases with decreasing Knudsen number. However, at $\mathrm{Kn}_\infty = 0.01$ (panel B, black curve) there are quite a few of them in this area, which is explained by more efficient charge exchange, and the atoms do not approach the star as closely. This indicates an expansion of the atom-free zone.

The dependence of Population 1 number density (panel A) with Knudsen number is essentially nonlinear. At first, with a decrease in the Knudsen number, the number density increases ($0.8 \leq \mathrm{Kn}_\infty \leq 24$), since the number of charge exchange events increases. Then it begins to decrease because the most of atoms do not reach the supersonic stellar wind, experiencing charge exchange in the inner astrosheath (and atom free-zone expands).

It should be noted that we were unable to reach a limit solution (the effective gas limit). In \ref{ap1}, the solution for $\mathrm{Kn}_\infty = 10^{-4}$ is discussed in detail. This is the lowest value of the Knudsen number that we have managed to achieve.

\subsection{Heated rarefied layer (HRL) in the outer astrosheath}
\label{PDL}

In this subsection, we discuss the effect of the displacement of the maximum of the number density in the outer shock layer away from the astropause towards the bow shock with, as the Knudsen number decreases (see Figure~\ref{fig1}, A, red curve, and diagram of the effect in Figure~\ref{shem1}, B). 

The plasma flow in the outer shock layer undergoes changes due to the momentum and energy source terms within the System of Equations (\ref{sys2}). Since four populations of atomic hydrogen are involved in the interaction with plasma, and the populations have different properties, understanding the roles of each population becomes a complex task. We conducted specific research within the framework of a toy model \citep{2024Korolkov} and explored the impact of various sources of momentum and energy upon the flow structure in a separate shock layer. The study allowed us to determine that heating of the shock layer leads to the displacement of the maximum plasma number density towards the shock wave. Conversely, cooling increases the number density near the astropause. Momentum sources directly influence the pressure distribution and the width of the shock layer, with almost no effect on the number density.

Applying the study mentioned above to our problem, we conclude that, for $\mathrm{Kn}_\infty\leq0.1$, the main effect is connected to hot atoms of the Population 2 described above (ENAs). Atoms of this population penetrate the outer shock layer and charge exchange with protons, providing a strong heat source. The greatest heating occurs at the astropause, leading to the redistribution of plasma density profiles and the formation of a Heated Rarefied Layer (HRL) of plasma in the outer astrosheath.

For larger Knudsen numbers, this effect does not completely disappear but is localized in a small layer near the astropause (see, for example, Figure~\ref{fig1}, A, $\mathrm{Kn}_\infty = 0.8$, blue curve, slight decrease in number density near the astropause at $X \approx 0.36$). Such a slight decrease in density is observed by Voyagers (see Figure 36, \cite{2022Richardson}), and It is called the Plasma Depletion Layer (PDL). Although the physical nature of the PDL is still unknown, we believe that the ENAs are partially responsible. This means that the PDL is (or is part of) a special case of the HRL described in this subsection.
 
Strong plasma density depletion towards the astropause has been obtained for $\mathrm{Kn}_\infty \lesssim 0.15$. These small Knudsen numbers correspond to astrospheres that are approximately $3$ times larger than heliosphere. Therefore, our modeling predicts a heated rarefied layer in the outer astrosheaths for stars with large astrospheres. If the other model parameters are similar to those in heliospheric, the mass loss rate of the star should be at least 9 times greater than solar one. In this scenario, the HRL would be very pronounced and could be observed.

Note that discussed in this subsection heated rarefied layer of plasma has not a one-dimensional nature. Displacement of the maximum of the number density in the outer shock layer occurs throughout the entire layer, even far from the axis of symmetry. Figure~\ref{pr1} in \ref{ap2} demonstrates this conclusion.


\subsection{Weakening of the bow shock}
\label{shockless}

\begin{figure}
	\includegraphics[width=1.0\hsize]{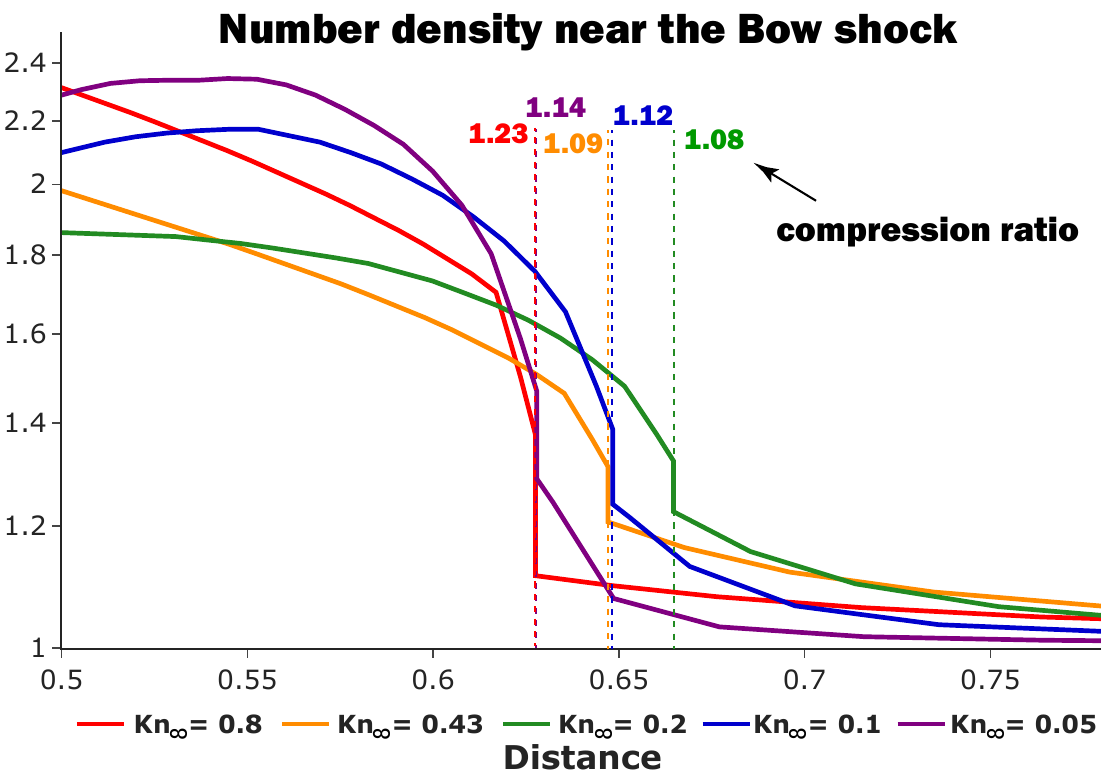}
	\caption{Dimensionles distributions of number density in the vicinity of bow shocks for various values of the Knudsen number. For each shock wave, its position (vertical lines) is marked, as well as the compression ratio - the ratio of the number density downstream and upstream the shock wave}
	\label{bowshock}
\end{figure}

In this subsection, we explore the nonlinear behavior of the bow shock with the Knudsen number within a range of $0.1 - 0.43$ (Figure~\ref{fig0}, highlighted in gray ellipse). The shock wave distance for this range is greater than the distance for $\mathrm{Kn}_\infty = 0.8$, although in all other ranges of the Knudsen number, it decreases monotonically with decreasing $\mathrm{Kn}_\infty$. 

Figure~\ref{bowshock} shows the number density distributions in the vicinity of the bow shock for various values of the Knudsen number in the range: [$0.05 - 0.8$]. The positions of the shock waves are additionally marked with vertical dotted lines. The compression ratio (the ratio of number densities downstream and upstream the shock wave) is marked with colored inscriptions for each wave. The intensity of bow shock is rather weak. For $\mathrm{Kn}_\infty = 0.2$ compression ratio is equal to $1.08$ and this corresponds to the minimum value that we found. It becomes even possible to speculate on a shockless transition. For values $\mathrm{Kn}_\infty = 0.1$ and $0.43$ the intensity is slightly higher, but still weak, so the position of the shock wave is further from the star. For Knudsen numbers outside this range, the intensity of shock waves increases again (see Figure~\ref{fig1}). The weakening of the bow shock is the reason for the slight movement of the shock wave further from the star in the $\mathrm{Kn}_\infty$ range from $0.1$ to $0.43$. \cite{Zank2013} showed the necessary condition for a shockless transition: 
\begin{align}
Q_3 = \dfrac{\gamma}{\gamma - 1} \bf{V} \cdot \bf{Q_2}. 
\end{align}
It is also fulfilled in our case, however, for other Knudsen numbers (for example, $\mathrm{Kn}_\infty = 3$) the expression $(Q_3 - \gamma/(\gamma - 1)\ \bf{V} \cdot \bf{Q_2})$ also changes the sign at the shock wave, although the shock wave is still present. Therefore, this condition is only necessary, but not sufficient. So far, it is only clear that the intensity of the wave decreases due to the influence of populations 1-3, which penetrate back into the interstellar medium and affect the incoming flow (this is so-called the \cite{1982Gruntman} effect). Note that for the solar value of $\mathrm{Kn}_\infty = 0.43$ the bow shock is also strongly weakened. Figure~\ref{pr1} in \ref{ap2} demonstrates this conclusion.


\section{Summary}
\label{C}

We have carried out a parametric study of the interaction of the stellar wind with the partly ionized interstellar medium, taking into account the charge exchange of protons on interstellar hydrogen atoms. The simulation was carried out in a wide range of Knudsen number parameters ($10^{-4} - 10^2$). Briefly, the results of the work can be summarized as follows:

(1) Without considering the influence of atoms ($\mathrm{Kn}_\infty \rightarrow \infty$, plasma-gas limit), the distances from the star to the discontinuity surfaces for $M_\infty = 1.97$ ($\gamma = 5/3$) can be calculated using Equation~\ref{LLL}. However, the lower the Knudsen number the more effective the charge exchange process, that leads to a decrease in the distances to discontinuity surfaces. Figure~\ref{fig0} shows the value of the coefficient $k$ (dimensionless distance) for various Knudsen numbers. It can be concluded that the maximal decreases of distances of the bow shock, astropause, and the termination shock compared to the plasma-gas limit are $\approx 36\%$, $\approx 47\%$, and $\approx 52\%$, respectively. Additionally, for astrospheres with $\mathrm{Kn}_\infty \geq 100$, the influence of the charge exchange process can be neglected.

(2) For Knudsen numbers $\mathrm{Kn}_{\infty} \leq 0.1$, intense localized heating occurs outside the astropause, due to heat flow from the interior of the AP, provided by charge exchange-created neutrals (ENAs) that flow across the AP and then charge exchange again. This causes the formation of the heated rarefied layer of plasma in the outer shock layer (see Figure~\ref{shem1}, B). We predict this phenomenon in stars with astrospheres three or more times larger than the heliosphere. In addition, the heated rarefied layer cannot be described in terms of single-fluid models, therefore it is ultimately necessary to employ a kinetic-gasdynamic approach for Knudsen number in the range of $10^{-4} < \mathrm{Kn}_\infty < 10^{2}$ (for smaller values of the Knudson number, these conclusions probably also remain valid). In addition to the kinetic approaches, multi-fluid approaches can also be used (see the discussion of approaches in Section~\ref{Intro}). However, the possibility of describing a heated rarefied layer in this case has not yet been studied.

(3) For the Knudsen number in the range of $0.1 - 0.5$ (including the heliospheric value $\approx 0.43$), we detect a weakening of the bow shock intensity. In this case, the position of weak shock wave is slightly greater ($\approx 6 \%$) than at $\mathrm{Kn}_\infty = 0.8$.




    

 Further research may be devoted to parametric studies of the influence of charge exchange on the structure of astrospheres by varying $\chi,\ M_\infty$, and the number density of interstellar hydrogen, as well as estimation of astrospheric parameters based on the visible positions of discontinuity surfaces and/or analysis of absorption profile of radiation from various stars.

\begin{acknowledgement}
The research is carried out using the equipment of the shared research facilities of HPC Resources of the Higher School of Economics \citep{Kostenetskiy_2021}.
\end{acknowledgement}



\paragraph{Competing Interests}

None

\paragraph{Data Availability Statement}

The data underlying this article will be shared on reasonable request to the corresponding author.

\printendnotes

\printbibliography

\appendix

\section{How far is the effective gas limit?}
\label{ap1}

\begin{figure*}
	\includegraphics[width=0.99\hsize]{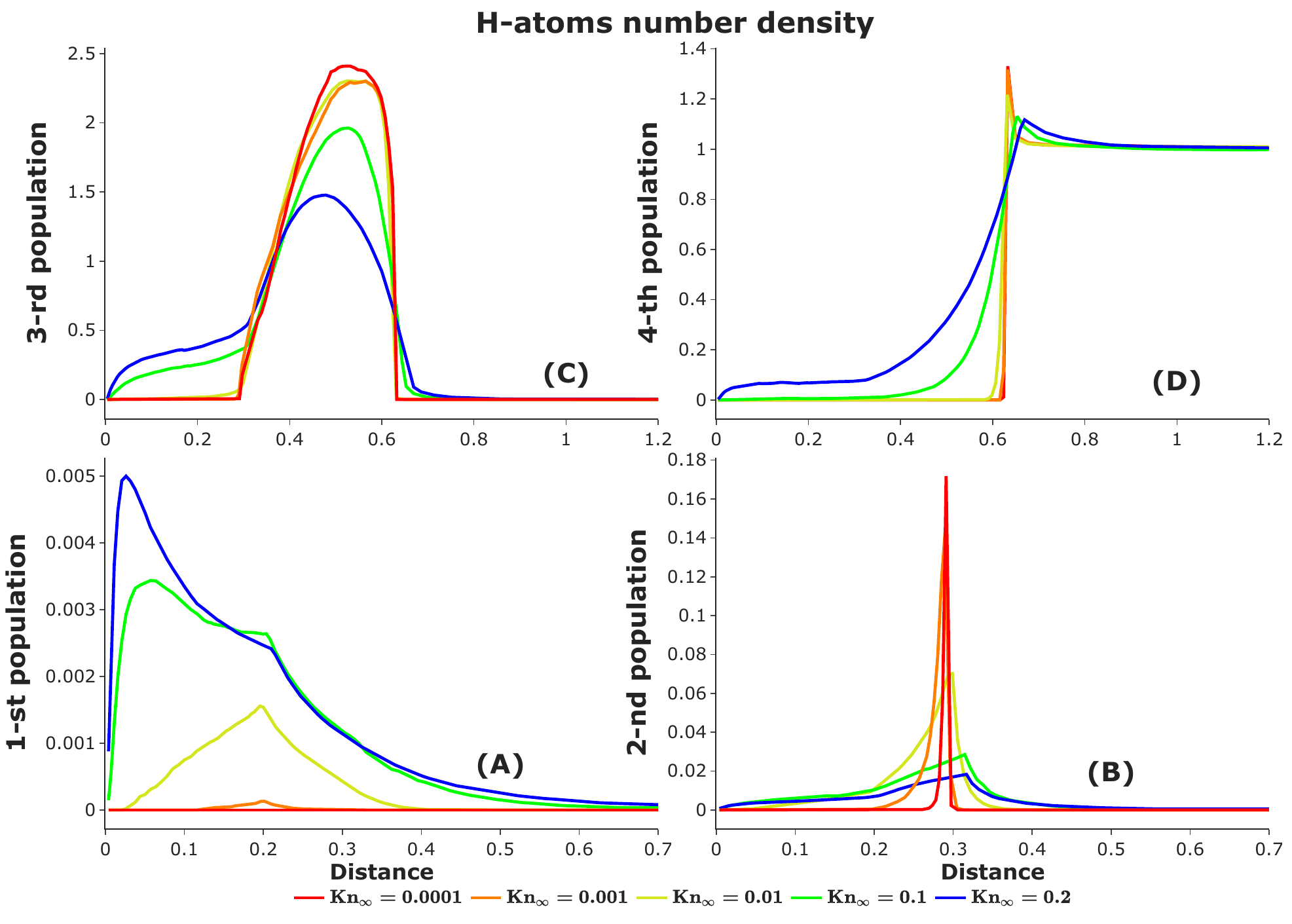}
	\caption{Number density of each population of hydrogen atoms for different values of the Knudsen number on the X-axis (upwind). Values are dimensionless for hydrogen concentration in LISM}
 \label{a1}
\end{figure*}

In this Appendix, we discuss the results of simulations with $\mathrm{Kn}_\infty \leq 0.2$, including the lowest Knudsen numbers that we were able to achieve ($\mathrm{Kn}_\infty = 10^{-4}$). It is also discussed how close we are to the effective gas limit.

Figure~\ref{a1} shows the distributions of the number density of Population 1 - 4 for different values of the Knudsen number at the X-axis (upwind) in the range of values: $10^{-4} \leq \mathrm{Kn}_\infty \leq 0.2$. Some values of the Knudsen number repeat Figure~\ref{fig2}. They are presented here just to compare.

The solution for $\mathrm{Kn}_\infty = 10^{-4}$ (Figure~\ref{a1}, red curve) is of greatest interest. In this case, there are no atoms of Population 1 (panel A, red curve). Moreover, the mean free path is so small that atoms of Population 2 are observed only in a narrow layer near the astropause (panel B, red curve, $X \approx 0.3$). This layer consists of atoms that are born as a result of the charge exchange of atoms of Population 3, which flew into the inner shock layer from the region located in proximity to the astropause in the outer astrosheath (in the vicinity of the stagnation point). Atoms of Population 3 (panel C) have parameters that are close to those of the plasma (the mean velocity and temperature are the same, and the number density is three times higher, since $\eta = 3$ for the selected parameters). For distributions of Population 4 (panel D, red curve), a thin layer of increased number density is observed near the bow shock (this is a precursor of the hydrogen wall, see Subsection~\ref{atom_par}). 

Although visually it seems (Figure~\ref{a1}, panel B, red curve) that there are almost no atoms of Population 2 in the outer shock layer, the sources (${\bf Q}_2,\ Q_3$) in the System~\ref{sys2} have a factor of $1/\mathrm{Kn}_\infty$, and therefore remain significant. This influence consists of heating the external heliosheath, and this heating is greater the closer the region is to the astropause. This heating is responsible for the formation of a heated rarefied layer of plasma in the outer astrosheath (see Subsection~\ref{PDL}). Thus, we assume that in the range $10^{-6} < \mathrm{Kn}_\infty < 10^{2}$ it is necessary to solve the Boltzmann kinetic equation to correctly describe the dynamics of hydrogen in astrospheres.


\section{2D plasma flow}
\label{ap2}

\begin{figure*}
	\includegraphics[width=0.99\hsize]{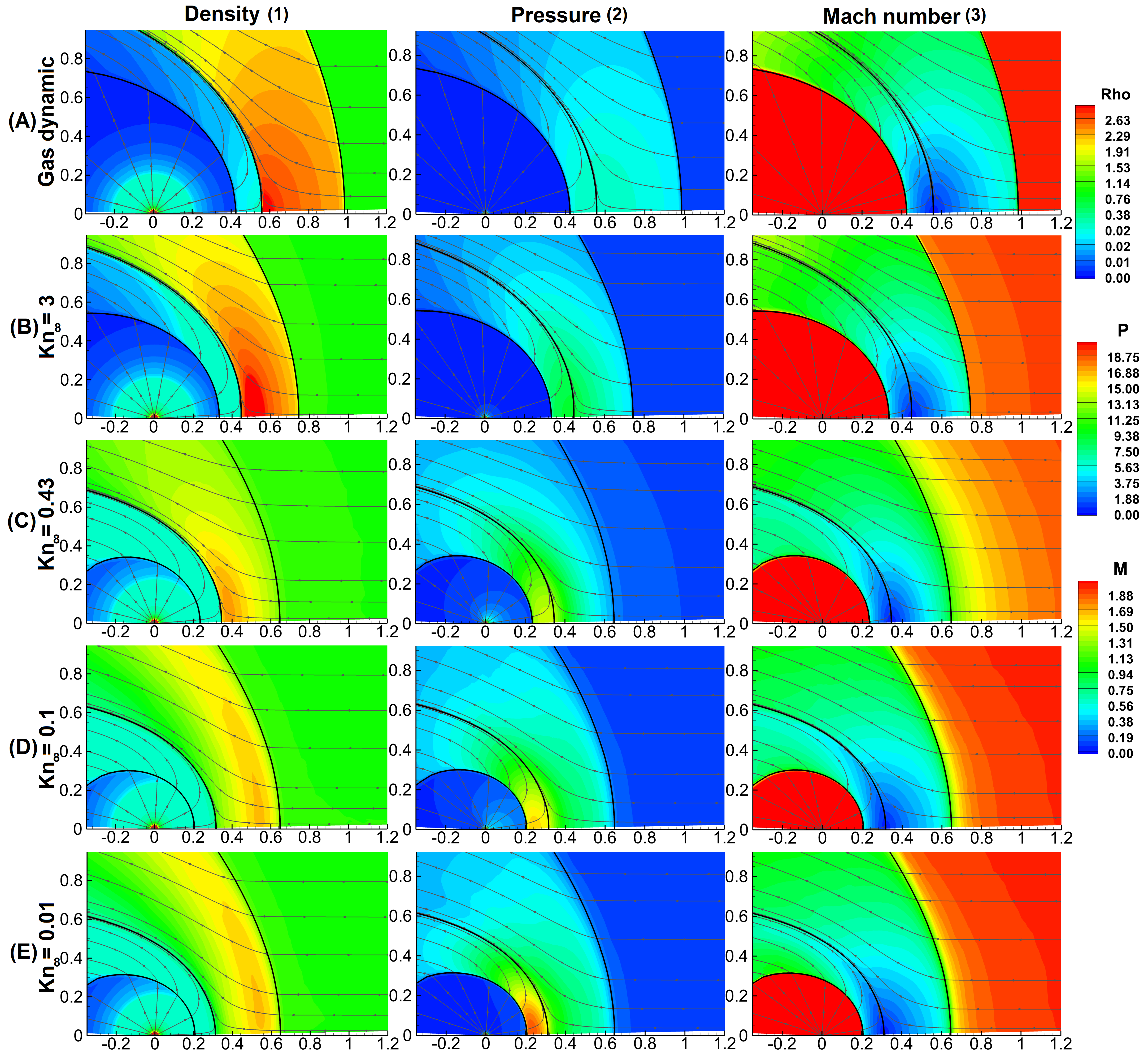}
	\caption{The streamlines and isolines of density (first column), pressure (second column), and the Mach number (third column) for different values of the Knudsen number. The density outflow to the bowshock is clearly visible for $\mathrm{Kn} = 0.01$. Density and pressure are dimensionless to the values in LISM}
	\label{pr1}
\end{figure*}

This Appendix provides 2D contour plots of the results from the simulations discussed in the main text.

Figure~\ref{pr1} presents 2-D flow patterns, number density (1), pressure (2), and Mach number (3) for different values of the Knudsen number. The color scale is specifically chosen to be the same for all panels. All panels have the same spatial range, making it easier to compare the astrospheric sizes. The color bars are uniform, which may slightly reduce the level of detail, but it does make it easier to compare and determine the magnitude of flow parameters. Additionally, the positions of the surfaces and streamlines are marked. The heated rarefied layer of plasma (see Subsection~\ref{PDL}) in the outer astrosheath is clearly visible at $\mathrm{Kn}_\infty \leq 0.1$ (1-D and 1-E, $0.3 \leq X \leq 0.65$). 

The weakening of the bow shock (see Subsection~\ref{shockless}) is clearly visible on the Mach number isolines for the heliospheric case ($\mathrm{Kn}_\infty = 0.43$, 3C). Here we can also determine the area of influence of atoms of populations 1-3 on the supersonic interstellar flow. It extends to values $X\approx 1$.

\end{document}